\documentstyle[11pt,newpasp,twoside,psfig]{article}
\index{summary}
\index{instructions}
\index{template}

\def\edcomment#1{\iffalse\marginpar{\raggedright\sl#1\/}\else\relax\fi}
\reversemarginpar

\def\ltsima{$\; \buildrel < \over \sim \;$}
\def\lsim{\lower.5ex\hbox{\ltsima}}
\def\gtsima{$\; \buildrel > \over \sim \;$}
\def\gsim{\lower.5ex\hbox{\gtsima}}

\def\axj {AX~J0851.9--4617.4~}

\def\sax  {SAX~J0852.0--4615~}

\begin{document}
\title{The X-ray sources at the center of   G266.1--1.2}
 \author{S. Mereghetti, A.Pellizzoni, A.De Luca}
\affil{Istituto di Fisica Cosmica ''G.Occhialini'', CNR, Milano}

\begin{abstract}
We present optical observations of the fields of two X--ray sources
located near the center of the shell-like supernova remnant G266.1--1.2.
No objects brighter than R$\sim$22.5 and B$\sim$23 are present within
the small \textit{Chandra} error region of \axj, besides a R$\sim$17 star that
has already been excluded as a possible counterpart.
A bright diffuse H$_{\alpha}$ nebula is present close to
the position of the candidate neutron star.
\end{abstract}

\section{Introduction}

The supernova remnant G266.1--1.2 has   been reported as a possible
$\gamma$-ray source in the 1.156 MeV
line of $^{44}$Ti (Iyudin et al. 1998).
The short lifetime ($\sim$90 yrs) of this isotope,  and
the relatively small angular size of the remnant would imply an age of
only $\sim$680 years and a small distance  d$\sim$200 pc (Aschenbach et al. 1999).
Thus G266.1--1.2 could be the remnant of the closest supernova event to have
occurred in recent historical times.

However, \textit{ASCA} observations showed that the X--rays from the SNR
shell have a non-thermal spectrum and the fits require a high
absorption value (Slane et al. 2001), favoring a distance of $\sim$1-2 kpc that would
place G266.1--1.2 well beyond the Vela SNR
(see also Mereghetti \& Pellizzoni 2001). The \textit{ASCA} data revealed
also a central point source, \axj , surrounded by diffuse X--ray
emission, that was interpreted as the neutron star associated to
G266.1--1.2.

A  \textit{BeppoSAX} observation (Mereghetti 2001) of the central
region of G266.1--1.2  showed the presence of a second source
about 3$'$ north of that detected by \textit{ASCA} and with a harder spectrum.
The northern source was named SAX~J0852.0--4615.
Since the \textit{BeppoSAX} error circle of \axj contained two
bright early type stars that might have produced the observed X--ray flux,
while no optical counterparts brighter than V$\sim$15 were visible for \sax,
it was unclear which of the two sources was the most likely neutron star candidate.

The puzzle has been recently solved by a \textit{Chandra} observation that provided an
arcsecond position for \axj (Pavlov et al. 2001).
The new error box is incompatible with  the two
early type stars that were previously considered as possible
counterparts, thus confirming that \axj is the most likely   neutron star
candidate.
\sax was not   detected in the 3 ks long
\textit{Chandra} observation reported by Pavlov et al. (2001).
This might be due to variability, or to the
hardness of this source, that was detected with \textit{BeppoSAX} only above 5 keV.
A deeper observation with \textit{XMM-Newton} confirmed the existence of \sax, with a
flux about ten times fainter than that of \axj (Aschenbach, this conference).

Here we present optical observations of the fields of these two X--ray sources.

\section{Optical observations of the field of \axj}

Optical images of the central region of G266.2--1.2 were
obtained through the public archive of the European Southern Observatory (ESO).
These data consist of images in the R and B bands taken with the
Wide Field Imager instrument at the La Silla 2.2m telescope on July 30, 1998.
We derived an approximate calibration of these data based on a number of
stars of the USNO catalogue.

Figure 1 shows the R band  image of the \textit{BeppoSAX}  error region
(1$'$ radius) of \axj.
The two early type stars that were previously considered as possible
counterparts  are the B8 type star HD76060 (V=7.9),
and Wray 16-30,  a  B[e] star with V=13.8  (Th\'{e} et al. 1994)
also detected with IRAS.

The small circle indicates the new position of \axj   determined with
\textit{Chandra} (Pavlov et al. 2001).
No objects, down to limits of R$\sim$22.5  and B$\sim$23 are present at
the \textit{Chandra} position,
besides a star with  R$\sim$17 on the edge of the error circle.
As discussed by Pavlov et al. this object is probably a late type star and
cannot be the counterpart of the X--ray source.

Some diffuse emission is also visible to the West of the \textit{Chandra} position.
This nebula is better seen in the H$_{\alpha}$ image shown in Fig.2.
The brightest part of the H$_{\alpha}$ nebula has a roughly elliptical shape with
dimensions $\sim$40$''$$\times$20$''$, and a sharply defined western edge.
Some nearly parallel filaments are also visible,
one of which seems to be composed by a series of aligned bright spots
(see Fig.1).

\begin{figure}
\centerline{\psfig{figure=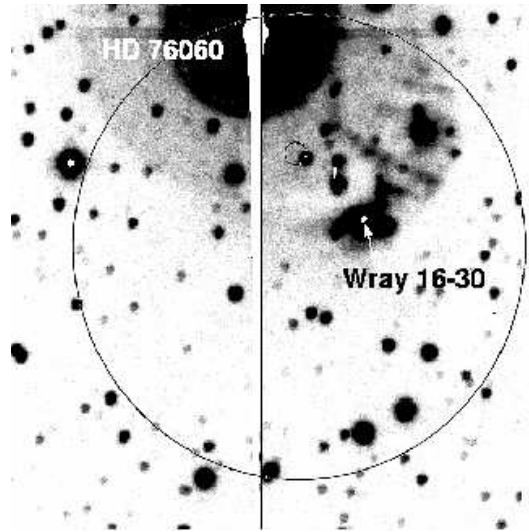,height=7cm,angle=0} }
\caption{Image in the R band of a 2$'$$\times$2$'$ region around the position
of \axj. The exposure time is 5 minutes. North is to the top, East to the left.
The circle with 1$'$ radius is the \textit{BeppoSAX} error region, while the
small circle indicates the \textit{Chandra} position derived by Pavlov et al. (2001). }
\label{f1}
\end{figure}

\begin{figure}
\centerline{\psfig{figure=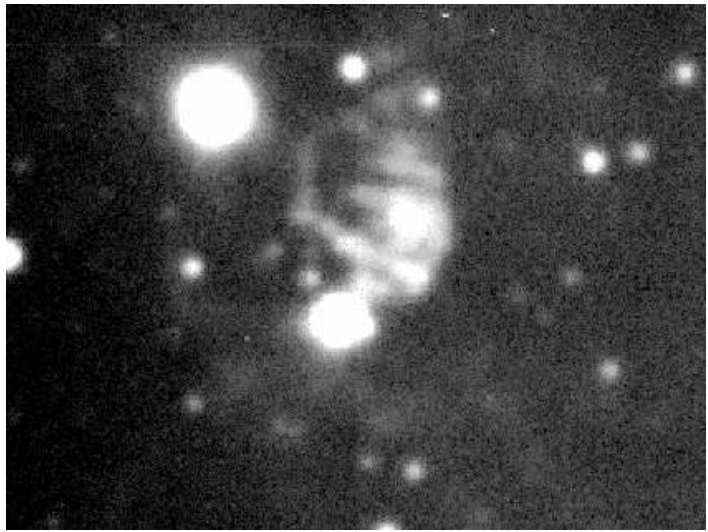,height=7cm,angle=00} }
\caption{The same region of Fig.1 imaged in  the H$_{\alpha}$ filter
with the EMMI instrument on the ESO NTT telescope  (exposure time 600 s).}
\label{f2}
\end{figure}

\section{Optical observations of the field of \sax}

The field of the northern source, \sax , is shown in the R band image of Fig.3.
The brightest stars in the \textit{BeppoSAX} error region (1$'$ radius) have R$\sim$15,
implying a ratio of X--ray to optical flux F$_x$/F$_{opt}$$>$0.1.
None of the stars in the error region has particularly unusual colors.

\begin{figure}
\centerline{\psfig{figure=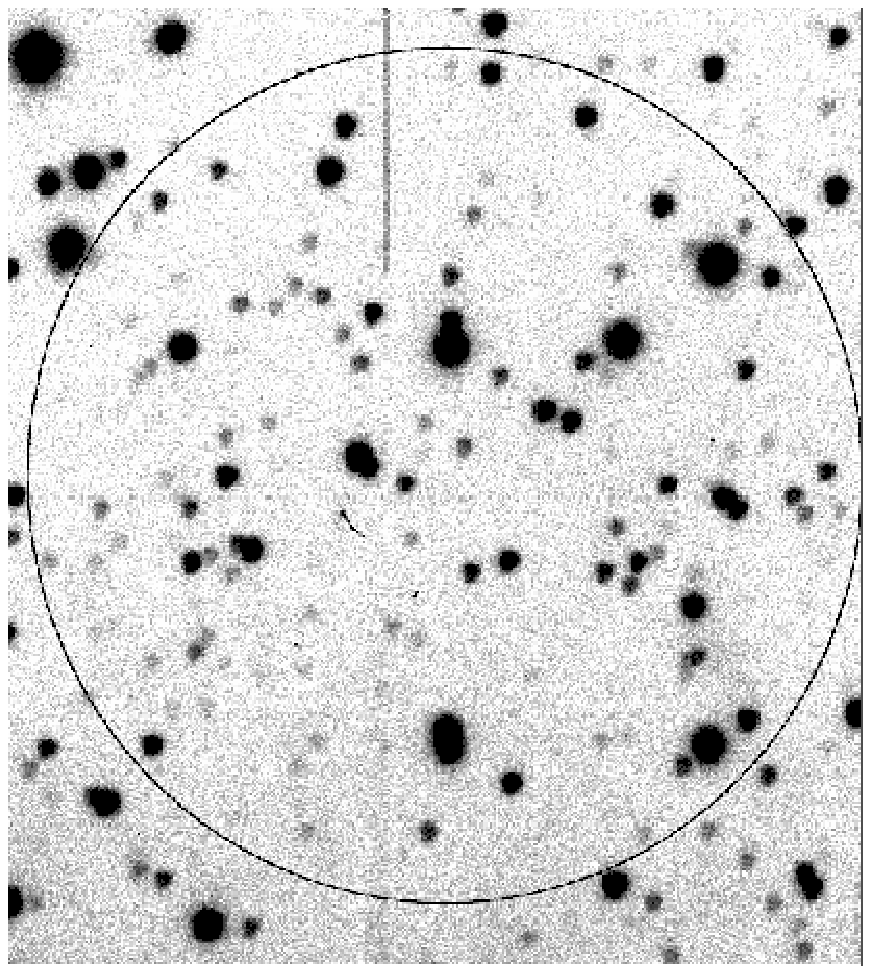,height=7cm,angle=0} }
\caption{Image in the R band of a 2$'$$\times$2$'$ region around the position
of \sax. The exposure time is 5 minutes. North is to the top, East to the left.
The circle with 1$'$ radius is the \textit{BeppoSAX} error region.}
\label{f3}
\end{figure}

\section{Conclusions}

The deep optical limits for the possible counterparts of \axj confirm
that this is most likely the neutron star remnant associated with G266.1--1.2.

An interesting H$_{\alpha}$ nebula has also been discovered in the data
presented here.
Emission in the H$_{\alpha}$ has been detected around a few radio pulsars
and is thought to originate in the interstellar medium shocked by the relativistic
pulsar wind. These nebulae have either a ''cometary'' shape  with the axis of symmetry
along the direction of the pulsar transverse motion (e.g.,
PSR B2224+65 (''Guitar Nebula'', Cordes et al. 1993) or PSR B0740--28 (Stappers et al.
these proceedings)) or an arc-like shape (e.g., PSR J0437--4715, Bell et al. 1996).

The morphology of the diffuse emission  shown in Fig.1 and Fig.2 does not
present any obvious connection with the location of the candidate neutron
star as determined with \textit{Chandra}. It is more likely that the nebula is   related to
the B[e] star Wray 16-30, which is located at the southern end of the nebula.
However, its peculiar morphology and the location close to the center of G266.1--1.2
make this nebula a potentially interesting target for more detailed investigations.

\acknowledgments

This work was based on data obtained with the ESO telescopes at La Silla and
made available through the ESO/STECF Science Archive Facility. We acknowledge
the support of the Italian Space Agency and of the Italian Ministry of
University and Research (Cofin-99-02-02).

\end{document}